# Critical Phenomena at the Antiferromagnetic Phase Transition of Azurite

P. T. Cong, B. Wolf, R. S. Manna, A. Brühl, S. Köhler and M. Lang

*Physikalisches Institut, J.W. Goethe-Universität Frankfurt, SFB/TR49, Frankfurt(M), Germany*

We report on high-resolution acoustic, specific-heat and thermal expansion measurements in the vicinity of the antiferromagnetic phase transition at $T_N$ = 1.88 K on a high-quality single crystal of the natural mineral azurite. A detailed investigation of the critical contribution to the various quantities at $T_N$ is presented. The set of critical exponents and amplitude ratios of the singular contributions above and below the transition indicate that the system can be reasonably well described by a three-dimensional Heisenberg antiferromagnet.

Email: cong@physik.uni-frankfurt.de

Fax: +49-69-798-47250

# I. INTRODUCTION

The natural mineral azurite, $Cu_3(CO_3)_2(OH)_2$, has been considered as a model substance for the 1D diamond chain, where anomalous magnetic properties result from the interplay of strong quantum fluctuations, low dimensionality and frustrating interactions [1, 2]. In a recent computational study a reasonable description was given for not too small energy scales (*i.e.*, $T > 2$ K and magnetic fields across the plateau region $B \leq 40$ T), covering a broad range of experimental results and placing azurite in the highly-frustrated parameter regime of the 1D distorted diamond chain [3]. However, early studies imply 3D long-range antiferromagnetic (AFM) order at $T_N \sim 1.88$ K, indicating that parameters such as inter-chain coupling and magnetic anisotropy play an important role and affect the magnetic properties of azurite in the low-temperature region [4]. The results of magnetic and structural investigations on azurite, recently performed at very low temperatures by means of neutron diffraction (ND) and muon-spin resonance (μSR) measurements, suggest a more complicated micromagnetic structure where magnetic order is accompanied by significant structural distortions [5]. In addition, the critical behavior of the AFM phase transition derived from these ND measurements yields an unusually small critical exponent (assigned to the order-parameter critical exponent $β$) of 0.06, indicating that the measured Bragg-peak intensity is not magnetic in origin. Therefore, in order to gain a better understanding of the magnetic transition and its critical behavior of azurite, we have conducted acoustic, specific heat and thermal expansion measurements around $T_N$. A detailed analysis of the critical behavior derived from these measurements has been performed and is summarized in this contribution.

# II. EXPERIMENTS AND RESULTS

The samples for the ultrasonic, specific heat and thermal expansion measurements were cut from a large high-quality single crystal that was also used in other measurements in refs. [2, 5]. The pair of thin-film transducers, used for the ultrasonic measurements, were glued to opposite polished parallel surfaces normal to the [010] crystallographic axis of the single crystal. These transducers generate longitudinal sound waves propagating along the [010] direction, corresponding to the acoustic $c_{22}$ mode. The velocity and attenuation of sound waves were measured using a phase-sensitive detection technique based on the standard ultrasonic pulse-echo method [6]. The measurements of the specific-heat, C($T$), were performed using a sensitive ac technique [7]. Measurements of the thermal expansion coefficient, $\alpha_L(T)$, have been carried out along the [010] axis employing a high-resolution capacitive dilatometer (built after [8]), enabling to detect length changes $\Delta l \geq 10^{-2}$ Å.

Figure 1 shows data of the acoustic $c_{22}$ mode (panel (*a*)), the specific heat (panel (*b*)) and the thermal expansion coefficient (inset of panel (*b*)) of azurite as a function of temperature close to the antiferromagnetic phase transition. The antiferromagnetic ordering at $T_N$ = 1.88 K is clearly visible as a sharp dip in the sound velocity and λ-like anomalies in the sound attenuation, specific heat and thermal expansion.

To check for the frequency dependence of the acoustic anomalies, the sound velocity and the attenuation were measured at different frequencies (not shown) ranging from 50 to 75 MHz, see [9] for more details. In the case of the sound velocity, there was no significant frequency dependence within the accuracy of the experiment, both above and below the transition. However, the attenuation increases by about a factor of two upon increasing the frequency from 50 MHz to 75 MHz. This frequency response is consistent with the results obtained in refs. [10, 11] indicating that the critical contribution to the sound velocity at a magnetic phase transition is frequency independent while that of the sound wave attenuation closely follows a quadratic frequency dependence. According to ref. 11, the critical contribution to the sound velocity $\Delta v_{crit}$ and attenuation $\alpha_{crit}$ can be described by power-law dependencies $-\frac{\Delta v_{crit}}{v_0} \sim \omega^0 \varepsilon^{-\varsigma}$ and $\alpha_{crit} \sim \omega^2 \varepsilon^{-\eta}$, respectively. Here $\zeta$ and $\eta$ are the critical exponents, $v_0$ is the velocity of the regular contribution at the critical point $T_N$, $\omega$ is the angular frequency of the acoustic mode and $\varepsilon$ is the reduced temperature $\varepsilon = (T - T_N)/T_N$. To extract the critical contribution, the non-critical background of the sound velocity and attenuation was subtracted by following the same procedure as outlined in [11].

Figure 2 shows the critical contribution to the sound attenuation (panel (*a*)) and sound velocity (panel (*b*)) deduced from Fig. 1(a) in a double-logarithmic plot. The main panels of (*a*) and (*b*) display the data for $T > T_N$ and the insets for $T < T_N$. Critical behavior can be clearly observed in the temperature range $\varepsilon > 6 \cdot 10^{-3}$ in the sound velocity and the sound attenuation both above and below $T_N$. The observed roll-off in the critical behavior very close to $T_N$ ($\varepsilon < 10^{-3}$) might be caused by a strong magnetoelastic-coupling [12] and/or by effects of impurities [13]. By fitting the attenuation data with a power law we find a critical exponent of $\eta \sim 1.12 \pm 0.002$ in the paramagnetic state ($T > T_N$) that is significantly different from $\eta \sim 5.42 \pm 0.2$, obtained in the ordered state ($T < T_N$). The value of $\eta = 1.12$ for the paramagnetic phase lies close to the theoretical prediction of $\eta = 1$ for the 3D isotropic antiferromagnet [14] but not too far from $\eta \sim 4/3$ for the 3D anisotropic antiferromagnet [15]. Here a perturbation treatment was made for the decay rate of one sound wave into two, three or four other sound waves and the scaling-law concept was used. The extraordinarily large value of $\eta$ in the ordered state might indicate additional extrinsic damping effects. As a possible source we mention the extraordinarily strong lattice distortion accompanying the magnetic transition which may lead to the formation of domains acting as scattering centres for the sound waves.

For the sound velocity, shown in Fig. 2(b), we obtain the same critical exponent of $\zeta \sim -0.056 \pm 0.001$ for $T > T_N$ and $T < T_N$. According to ref. 11 the critical contribution of the sound velocity is expected to show the same power-law behavior as the specific heat, *i.e.*, $\zeta \approx \alpha$. Therefore $\zeta \sim -0.056$ is indicative of a 3D isotropic antiferromagnet, for which $\alpha = -0.12$ is predicted [16], rather than the anisotropic variant where $0 \leq \alpha \leq 0.14$ [16].

To determine the critical exponent of the specific heat $C(T)$, the data in the vicinity of $T_N$ are commonly evaluated with the function (1) to describe both the critical and non-critical part [17].

$$C^{\pm} = (A^{\pm}/\alpha)|\varepsilon|^{-\alpha} + B + E\varepsilon \tag{1}$$

The first term in eq. (1) represents the leading contribution to the singularity in $C(T)$ with amplitudes $A^{\pm}$, where the superscripts + and − in $C$ and $A$ correspond to $\varepsilon > 0$ and $\varepsilon < 0$, respectively. The regular contributions are approximated by a linear $\varepsilon$ dependence, which is known to be sufficient to describe the data in the restricted fit range near $T_c$ adequately. In the present case, the fit range was limited to the interval $0.01 < \varepsilon < 0.1$ that is applied to the acoustic data in order to exclude the data very close to $T_c$ where roll-off effects are visible (see Fig. 3) and to eliminate the data outside the critical region. The thermal expansion coefficient $\alpha_L(T)$ is evaluated by using the same function (1) as described above. The resulting fitting curves for the specific heat are shown as solid lines in Fig. 3(a), where $C$ is plotted against $\log(\varepsilon)$. The fit gives $\alpha = -0.0912 \pm 0.002$ and an amplitude ratio $A_C^+/A_C^- = 1.32(1)$. This result lies closer to the theoretical predictions for the 3D Heisenberg afm, yielding $\alpha = -0.11$ and $A^+/A^- \sim 1.5$ [16], rather than the anisotropic case where $0 \leq \alpha \leq 0.14$ and $0.54 \leq A^+/A^- \leq 1$ [16].

The same fitting procedure has been applied to the thermal expansion data shown in Fig. 3(b). The fits (solid lines in Fig. 3(b)) yield a critical exponent of $-0.0853 \pm 0.002$ and $A_L^+/A_L^- = 1.02(1)$, in good agreement with the specific heat result and in reasonable agreement with the predictions for a 3D isotropic Heisenberg antiferromagnet.

### III. CONCLUSION

High-resolution acoustic, specific heat and thermal expansion measurements have been performed around the antiferromagnetic transition of azurite. The frequency dependence observed in the acoustic measurements, yielding $\alpha \propto \omega^2$ and $\frac{\Delta v_{crit}}{v_0} \propto \omega^0$, is consistent with the predictions for an antiferromagnetic transition [11]. The set of critical exponents derived from the various quantities, i.e., $\eta \sim 1.12$, $\zeta \sim -0.056$ and $-0.092 \leq \alpha \leq -0.085$, lie close to the predictions for a 3D Heisenberg antiferromagnet. Significant deviations from this behavior are observed only for the ultrasonic attenuation below the transition. We speculate that this effect is of extrinsic nature and related to the formation of domains as a consequence of the extraordinarily strong lattice deformations which accompany the magnetic transition. Note that for the 3D Heisenberg antiferromagnet an order parameter critical exponent $\beta = 0.36$ is expected. This is significantly different from $\beta = 0.06$ derived from neutron diffraction experiments for the present material [2]. The reason for this discrepancy is unclear at present and deserves further investigations.

Figure Captions.

Fig. 1. a) Temperature dependence of the acoustic $c_{22}$ mode (sound velocity on the right scale and relative change of the sound attenuation on the left scale) around $T_N$ of azurite. b) Specific heat C(T) (main panel) and the linear thermal-expansion coefficient $\alpha_L$(T) (inset) vs. T around $T_N$ of azurite.

Fig. 2. Double-log representation of the sound attenuation (shown in panel (a), and the sound velocity, (shown in panel (b)) for 75 MHz longitudinal waves vs. reduced temperature ε for $T > T_N$ (main panel) and $T < T_N$ (inset).

Fig. 3. The specific heat (panel (a)) and thermal expansion coefficient (panel (b)) vs. reduced temperature ε for $T > T_N$ (open squares) and $T < T_N$ (open circles). Solid lines represent the fitting curves.

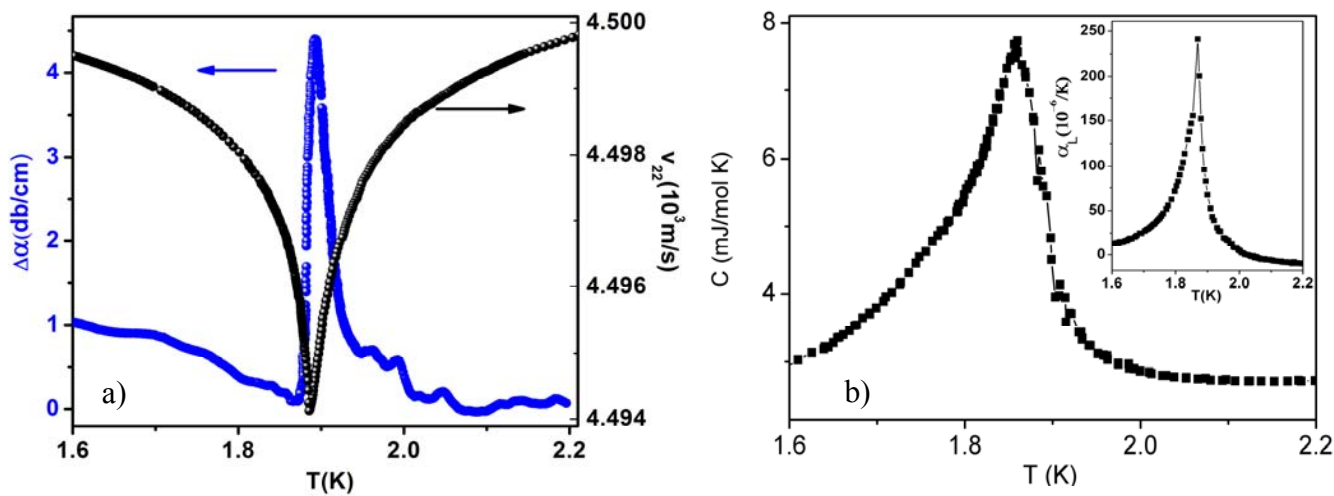

Fig.1

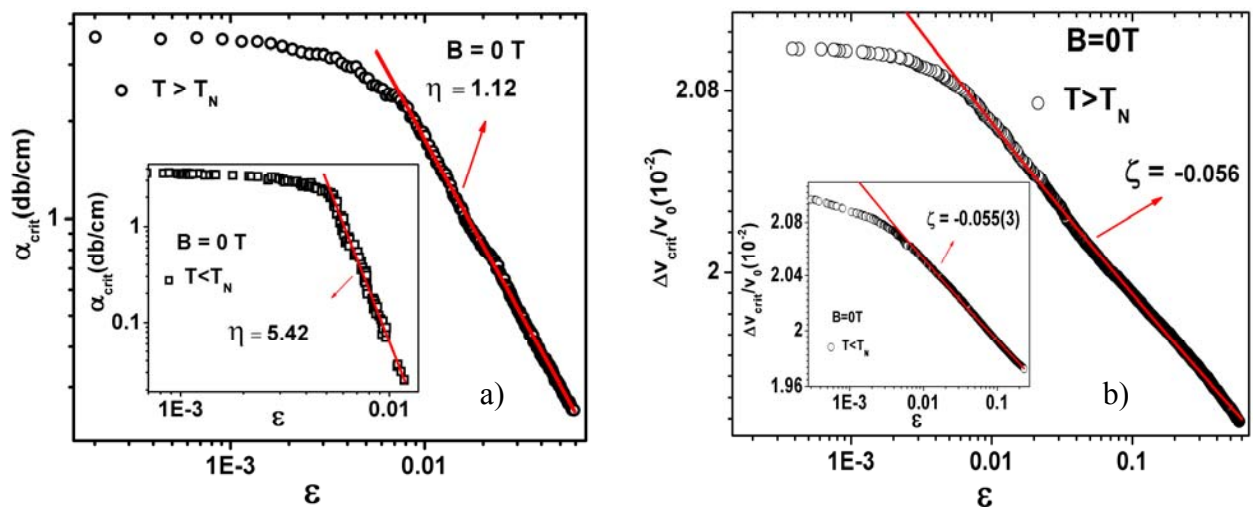

Fig. 2

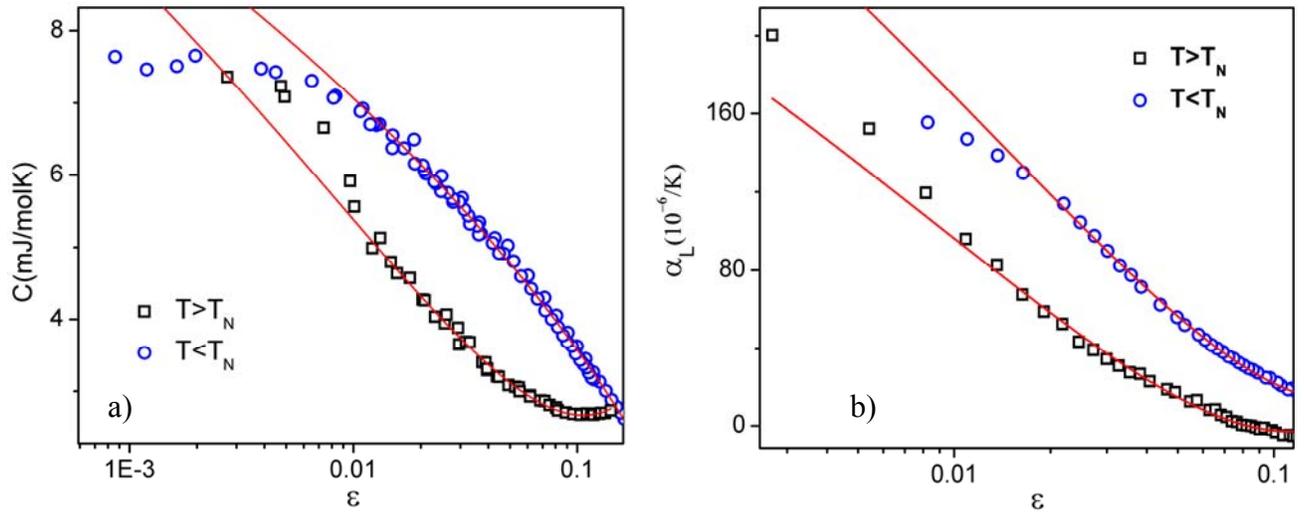

Fig. 3